\documentclass{article}
\usepackage{spconf,amsmath,graphicx}
\usepackage{cite,bm}
\usepackage{amssymb}
\usepackage{amsfonts}
\usepackage{comment}
\usepackage{amsmath,mathtools}
\usepackage{booktabs}
\usepackage{multirow}
\usepackage[binary-units]{siunitx}
\usepackage{xcolor}
\usepackage{physics}
\AtBeginDocument{
  \abovedisplayskip     =0.25\abovedisplayskip
  \abovedisplayshortskip=0.25\abovedisplayshortskip
  \belowdisplayskip     =0.25\belowdisplayskip
  \belowdisplayshortskip=0.25\belowdisplayshortskip
}

\def\L{{\mathcal L}}
\def\T{{\mathcal T}}
\def\S{{\mathcal S}}

\def\R{{\mathbb R}}
  
\usepackage{arydshln}
\usepackage{url}
\ninept

\def\L{{\cal L}}

\title{Self-Remixing: Unsupervised Speech Separation\\via Separation and Remixing}

\name{
    Kohei Saijo,
    Tetsuji Ogawa
}
    
\address{
Department of Communications and Computer Engineering, Waseda University, Tokyo, Japan
}

\begin{document}
\maketitle

\begin{abstract}
\vspace{-1mm}

We present \textit{Self-Remixing}, a novel self-supervised speech separation method, which refines a pre-trained separation model in an unsupervised manner.
Self-Remixing consists of a \textit{shuffler} module and a \textit{solver} module, and they grow together through separation and remixing processes.
Specifically, the shuffler first separates observed mixtures and makes pseudo-mixtures by shuffling and remixing the separated signals.
The solver then separates the pseudo-mixtures and remixes the separated signals back to the observed mixtures.
The solver is trained using the observed mixtures as supervision, while the shuffler's weights are updated by taking the moving average with the solver's, generating the pseudo-mixtures with fewer distortions.
Our experiments demonstrate that Self-Remixing gives better performance over existing remixing-based self-supervised methods with the same or less training costs under unsupervised setup.
Self-Remixing also outperforms baselines in semi-supervised domain adaptation, showing effectiveness in multiple setups.

\vspace{-1mm}

\end{abstract}
\vspace{-2mm}
\section{Introduction}
\label{sec:intro}
\vspace{-2mm}

Speech separation has been used as a front-end of voice applications such as automatic speech recognition (ASR).
Neural networks have led to remarkable progress in speech separation, driven mainly by supervised learning with a large number of pairs of mixtures and ground truths~\cite{pit}, synthesized using room impulse responses (RIRs) obtained by simulation toolkits~\cite{pra}.
While models trained on such data perform well in matched conditions, they perform poorly on actual recordings due to the mismatch with simulated conditions.

Such a mismatch can be mitigated with unsupervised learning, where models are trained with unlabelled data from the target domain~\cite{unsupervised_dc, drude_unsup, togami_unsup, spatial_loss}.
Mixture invariant training (MixIT)~\cite{mixit} has made unsupervised learning on monoaural mixtures possible by training models to separate a mixture of mixtures (MoM).
Although MixIT has been shown to work on a variety of separation tasks~\cite{efficient_mixit, bird_mixit, adapting_mixit}, MoMs cause the mismatch between training and inference since they unduly contain more sources than individual mixtures, resulting in over-separation problem and limiting the performance~\cite{tsmixit, samixit}.

\begin{figure}[t!]
\centering
\centerline{\includegraphics[width=0.8\linewidth]{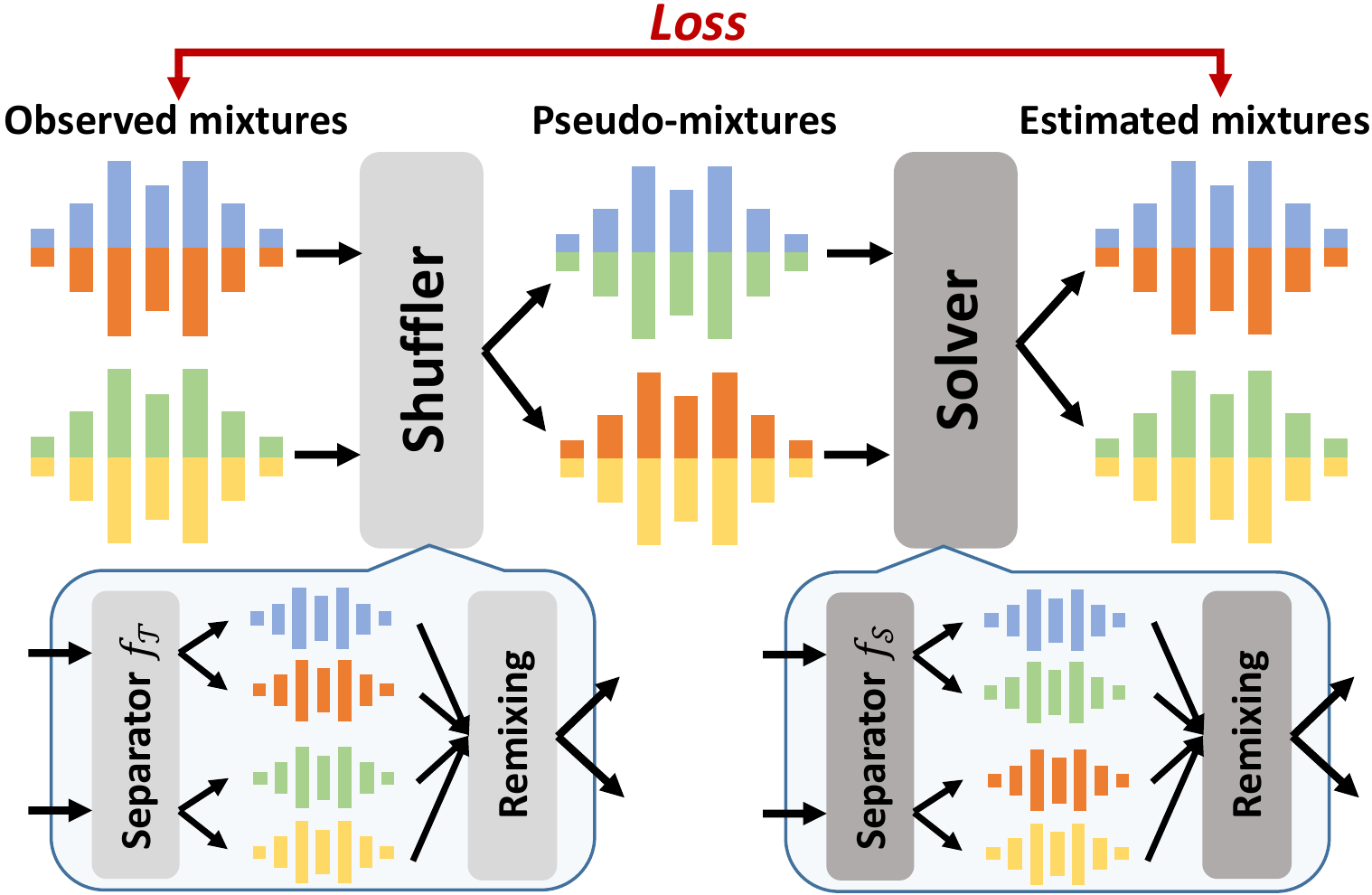}}
\vspace{-2.5mm}
\caption{
    Conceptual image of proposed Self-Remixing.
    Shuffler first makes pseudo-mixtures by remixing its separation outputs (\textit{making puzzle}), and solver recovers observed mixtures from pseudo-mixtures by separation and remixing (\textit{solving puzzle}).
    Note that Self-Remixing can be applicable to mixtures with any number of sources.
}
\label{fig:conceptual_image}
\vspace{-2mm}
\end{figure}

In this paper, we propose a novel self-supervised learning framework, \textit{Self-Remixing}, which can refine pre-trained separation models in an unsupervised manner (Fig.~\ref{fig:conceptual_image}).
Unlike MixIT, Self-Remixing does not change the number of sources in inputs, and thus alleviates the MixIT's mismatch problem.
Self-Remixing grows two modules together and can be likened to a relationship between a \textit{shuffler} that makes a puzzle and a \textit{solver} that solves it.
The solver can learn how to solve the puzzle by seeing the original one before broken up, while the shuffler can make a better puzzle by learning how the solver solves it.
In Self-Remixing, the shuffler first separates observed mixtures and makes pseudo-mixtures (puzzles) by remixing the separated signals in different combinations than the original.
The solver then separates the pseudo-mixtures and remixes the separated signals back to the observed mixtures (solves the puzzles).
While the separation performance of the solver gets better, aiming to reconstruct the observed mixtures from the pseudo-mixtures, the shuffler's weights can also be updated using the solver's, generating better pseudo-mixtures.
We efficiently refine the shuffler by taking the moving average of the parameters of the two modules.

The proposed method is related to recently proposed two self-supervised methods, RemixIT~\cite{remixit, remixit_journal, mixpit} and remix-cycle-consistent learning (RCCL)~\cite{hoshen_rccl, saijo_rccl, saijo_rccl2}.
RemixIT is a self-training method where a student model is trained to estimate the teacher's outputs by taking pseudo-mixtures generated with the teacher's outputs as inputs.
While the shuffler in Self-Remixing only gives pseudo-mixtures and teaches nothing to the solver, the RemixIT's teacher provides both pseudo-mixtures and supervision to the student.
Since RemixIT heavily relies on the teacher, it is expected to be unstable when the teacher is frequently updated.
In contrast, RCCL aims to recover observed mixtures through separation and remixing processes like the proposed method, but the same separator is used in the shuffler and the solver.
Compared with Self-Remixing, RCCL is unstable and likely to fall into the trivial solution, i.e., models do not separate inputs at all because the same module makes pseudo-mixtures and separates them.
In addition, RCCL computes gradients through two separation processes, which is problematic in terms of memory cost.
Our experiments demonstrate the superiority of the proposed method in terms of performance, stability, and efficiency in unsupervised speech separation and semi-supervised domain adaptation.

The key contributions are summarized as follows.
1)~We present a novel self-supervised speech separation framework.
We show the effectiveness of the proposed method in challenging unsupervised setups and more realistic semi-supervised setups.
2)~We conduct experiments using both synthetic and recorded data, whereas most prior work of single-channel separation only considered the former.
3)~This is the first work to apply RemixIT and RCCL for MixIT or PIT pre-trained model in source separation tasks.
Although we evaluate them as baselines, we introduce several new techniques to make them work well, including remixing method and loss thresholding.
\vspace{-2mm}
\section{Problem Formulation}
\label{sec:problem_formulation}
\vspace{-1.5mm}

Let us denote a mini-batch of $B$ microphone observations as $\bm{x}\in\mathbb{R}^{B \times T}$, with $T$ samples in the time domain.
Each mixture in the mini-batch, $x_b (b=1,\dots,B)$, consists of up to $K$ sources.
We define a separator with $N_\S$ output channels as $f_\S$, with parameters $\bm{\theta}_\S$.
In the following, we describe methods used for pre-training.
\\
\textbf{Permutation Invariant Training (PIT):}
Given the reference signal for the $n$-th source of the $b$-th mixture, $s_{b,n}$, and the separated signals $\hat{\bm{s}}_{b}=f_\S(x_b;\bm{\theta}_\S)\in\mathbb{R}^{N_\S \times T}$, PIT loss is computed as~\cite{pit}:
\begin{align}
  \label{eqn:pit_loss}
    \L_{\rm{PIT}}^{(b)} = \min_{\bm{P}} \sum\nolimits_{n=1}^{N_\S} {\L(s_{b,n}, [\bm{P}_{b}\hat{\bm{s}}_{b}]_{n})}, 
\end{align}
where $\bm{P}$ is a permutation matrix and $\L$ is a loss function.

\noindent\textbf{Mixture Invariant Training (MixIT):}
Given two mixtures $x_1$ and $x_2$ from the mini-batch $\bm{x}$, MixIT generates a mixture of mixtures (MoM) $\bar{x}$ by adding them together, $\bar{x}=x_1+x_2$.
MixIT loss is computed between the individual mixtures and the separated signals from the MoM, $\hat{\bm{s}} = f_\S(\bar{x};\bm{\theta}_\S)\in\R^{N_\S \times T}$, (${N_\S} \geq 2K$) as~\cite{mixit}:
\begin{align}
  \label{eqn:mixit_loss}
    \L_{\rm{MixIT}} = \min_{\bm{A}} \sum\nolimits_{i=1}^{2} {\L(x_i, [\bm{A}\hat{\bm{s}}]_{i})},
\end{align}
where $\bm{A}$ is a $2 \times N_\S$ matrix that assigns each $\hat{s}_{n}$ to either $x_1$ or $x_2$.
Although MixIT has made unsupervised monaural source separation possible, MoMs contain more sources than individual mixtures, causing the over-separation problem and limiting the performance.
\vspace{-2mm}
\section{Proposed Self-Remixing Method}
\label{sec:methods}
\vspace{-2mm}

\subsection{Basic framework}
\label{ssec:basic_framework}
\vspace{-1.5mm}

The proposed method consists of a shuffler that creates pseudo-mixtures from observed mixtures, and a solver that recovers the observed mixtures from the pseudo-mixtures (Fig.~\ref{fig:conceptual_image}).
In addition to $f_\S$ corresponding to the solver's separator, we define shuffler's separator with $N_\T$ outputs as $f_\T$, with the parameters $\bm{\theta}_\T$.
We assume $f_\T$ is pre-trained by MixIT with in-domain data or PIT with out-of-domain (OOD) data.
Although Self-Remixing works regardless of the number of sources, we assume each mixture contains up to two speech sources and a noise source, i.e., $K=3$.

Given mixtures $\bm{x}$, we first separate them with the shuffler's separator: $\tilde{\bm{s}}=f_\T(\bm{x};\bm{\theta}_{\T})\in\mathbb{R}^{B \times N_\T \times T}$.
We then select $N_\mathcal{R}$ sources with the highest powers in each $\tilde{\bm{s}}_{b}$.
When using MixIT pre-trained model, for example, the model has more outputs than the number of sources in mixtures (i.e., $N_\T \geq 2K$).
We aim to alleviate MixIT's over-separation problem by remixing only less than $N_\T$ sources. 
$N_\mathcal{R}$ is empirically determined depending on the remixing method, described in detail later.
After the selection, we enforce mixture consistency (MC)~\cite{mixconsis} to ensure the sources to add up to the observed mixtures, i.e., $\sum\nolimits_{n=1}^{N_\mathcal{R}}{\tilde{s}_{b,n}} = x_b$.
Next, we shuffle the sources and obtain pseudo-mixtures $\tilde{\bm{x}}$ by adding them.
Finally, the solver separates the pseudo-mixtures: $\hat{\bm{s}}=f_\S(\tilde{\bm{x}};\bm{\theta}_{\S})\in\mathbb{R}^{B \times N_\S \times T}$,
and $N_\mathcal{R}$ sources are selected in the same way as shuffler's outputs and used to compute the loss.
Note that only the solver $f_\S$ is trained with backpropagation, and the shuffler's weights $\bm{\theta}_\T$ are updated using the solver's weights $\bm{\theta}_\S$.
Specifically, we update $\bm{\theta}_\T$ via the exponential moving average update for efficient training (see Sect.~\ref{ssec:shuffler_update}).
In the following, we describe two remixing methods used in this paper, remixing between two mixtures and remixing in a batch.

\vspace{-2mm}
\subsubsection{Remixing between two mixtures}
\label{sssec:between_two_mixtures}
\vspace{-1.5mm}

In this method, the sources separated from \textit{two} mixtures are remixed as in RCCL~\cite{saijo_rccl}.
Given $\bm{s}_1$ and $\bm{s}_2 \in \R^{N_\T \times T}$, the sources of $x_1$ and $x_2$ separated by $f_\T$, we remix all $N_\T$ sources (i.e., $N_\mathcal{R}=N_\T$):
\begin{align}
    \label{eqn:rccl_remix}
    \tilde{x}_{1} = {\bm{\pi}}_{1}\tilde{\bm{s}}_{1} + (\bm{1}-{\bm{\pi}}_{2})\tilde{\bm{s}}_{2},~~~
    \tilde{x}_{2} = (\bm{1} - {\bm{\pi}}_{1})\tilde{\bm{s}}_{1} + {\bm{\pi}}_{2}\tilde{\bm{s}}_{2},
\end{align}
where the $1 \times N_\mathcal{R}$ vector ${\bm{\pi}}$ assigns sources to either $\tilde{x}_{1}$ or $\tilde{x}_{2}$.
Here, we assign the source with the highest power in each of $\tilde{\bm{s}}_1$ and $\tilde{\bm{s}}_2$ to $\tilde{x}_1$ and that with the second highest power to $\tilde{x}_2$, respectively.
The other $N_\mathcal{R}-2$ sources are assigned randomly under the condition of assigning the same number of sources to $\tilde{x}_{1}$ and $\tilde{x}_{2}$.
After $f_\S$ separates the pseudo-mixtures, $\tilde{x}_{1}$ and $\tilde{x}_{2}$, we compute the loss as:
\begin{align}
  \label{eqn:rccl_loss}
    \L_{\rm{Prop1}} = \min_{\bm{p}} ({\L(x_1, \bm{p}_{1}\hat{\bm{s}}_{1}+(\bm{1}-\bm{p}_{2})\hat{\bm{s}}_{2})}\;\;\;\;\;\; \nonumber\\
    \;\;\;\;\;\;+ {\L(x_2, (\bm{1} - \bm{p}_{1})\hat{\bm{s}}_{1}+\bm{p}_{2}\hat{\bm{s}}_{2})}),
\end{align}
where $\bm{p}$ is a $1 \times N_\mathcal{R}$ permutation vector that assigns sources to either $x_{1}$ or $x_{2}$.
The optimal permutation is obtained with an exhaustive $\mathcal{O}((C(N_{\mathcal{R}}, N_{\mathcal{R}}/2))^2)$ search, where $C(u,v) = u! / (v!(u-v)!)$.

\vspace{-2.5mm}
\subsubsection{Remixing in batch}
\label{sssec:in_batch}
\vspace{-1.5mm}

In this method, the sources are shuffled in a mini-batch, and the training process of Self-Remixing follows RemixIT except for the loss computation.
Given the mini-batch of sources, $\tilde{\bm{s}} \in \R^{B \times N_\T \times T}$, we select three sources with the highest powers ($N_\mathcal{R}=3$) and remix:
\begin{align}
  \label{eqn:remixit_remix}
    \tilde{x}_{b} = \sum\nolimits_{n=1}^{N} {\tilde{s}_{b,n}}^{(\bm{\Pi})}, ~~~~~~~~{\tilde{s}_{b,n}}^{(\bm{\Pi})} = [{{\bm{\Pi}}_n}{{\tilde{\bm{s}}}^{\top}_n}]_{b},
\end{align}
where $^{\top}$ denotes the transpose of the first and the second dimension ($\R^{B \times N_\mathcal{R} \times T} \rightarrow \R^{N_\mathcal{R} \times B \times T}$) and ${\tilde{\bm{s}}}^{\top}_n \triangleq [\tilde{s}_{1,n}, \ldots, \tilde{s}_{B,n}] \in \R^{B \times T}$.
${{\bm{\Pi}}_n}$ is a $B \times B$ permutation matrix that shuffles sources within each output channel and is chosen under the condition that sources separated from the same mixture are not remixed again.
After $f_\S$ separates $\tilde{\bm{x}}$, we select three sources.
Here, when using MixIT pre-trained model, we make up for the discarded sources with MC, $\sum\nolimits_{n}{\hat{s}_{b,n}} = \tilde{x}_b$.
We then search the permutation matrix $\bar{\bm{P}}$ that minimizes the following cost function to align the outputs of $f_\T$ and $f_\S$:
\begin{align}
  \label{eqn:remixit_loss}
    \L_{\rm{RemixIT}}^{(b)} = \min_{\bm{P}} \sum\nolimits_{n=1}^{N_\mathcal{R}} {\L({\tilde{s}}^{({\bm{\Pi}})}_{b,n}, [\bm{P}_{b}\hat{\bm{s}}_{b}]_{n})}.
\end{align}
Note that Eq.~(\ref{eqn:remixit_loss}) is equal to the RemixIT loss~\cite{remixit_journal}.
In contrast to RemixIT which trains $f_\S$ using the outputs of $f_\T$ as supervision, Self-Remixing uses the observed mixture $x_b$ as supervision:
\begin{align}
  \label{eqn:proposed_loss}
    \L_{\rm{Prop2}}^{(b)} = \L\left(x_b, \sum\nolimits_{n} {{\hat{s}_{b,n}}^{(\bm{\Pi}^{-1})}}\right),~~
    {{\hat{s}_{b,n}}^{(\bm{\Pi}^{-1})}} = [{{\bm{\Pi}}^{-1}_n}{{\hat{\bm{s}}}^{(\bar{\bm{P}})\top}_n}]_{b},
\end{align}
where ${\hat{\bm{s}}^{(\bar{\bm{P}})}}_{b} = \bar{\bm{P}}_{b}\hat{\bm{s}}_{b}$ denote the solver's outputs aligned with shuffler's and ${{\hat{\bm{s}}}^{(\bar{\bm{P}})\top}_n} \triangleq [\hat{s}^{(\bar{\bm{P}})}_{1,n}, \ldots, \hat{s}^{(\bar{\bm{P}})}_{B,n}] \in \R^{B \times T}$.
Multiplying $\bm{\Pi}^{-1}$ restores the original order of sources before shuffled in Eq.~(\ref{eqn:remixit_remix}).
Computing Eq.~(\ref{eqn:proposed_loss}) needs $\mathcal{O}(N_\mathcal{R}!)$ search to find $\bar{\bm{P}}$.
Note that we do compute the RemixIT loss to compute the Self-Remixing loss efficiently, but we use only the latter for updating the model parameters.

\vspace{-2mm}
\subsection{Loss thresholding}
\label{ssec:loss_thresholding}
\vspace{-1.5mm}

In preliminary experiments, we found that the separation performance improved in the early stage of training but deteriorated after the loss dropped to a certain level when remixing between two mixtures.
We believe this is because both perfect separation and not separating at all minimize Eq.~(\ref{eqn:rccl_loss}).
Specifically, while models first try to reduce the loss by producing better-separated signals, they come to output the mixture itself to further reduce the loss after reaching the limit of performance improvement.
To mitigate this problem, we introduce simple thresholding where we multiply the loss by zero when Eq.~(\ref{eqn:rccl_loss}) is less than $l_{\rm{thres}}$.
This prevents the loss from dropping too low and falling into the trivial solution.
Note that we empirically confirm that thresholding is not necessary when remixing in a batch.

\vspace{-2mm}
\subsection{Shuffler update protocol}
\label{ssec:shuffler_update}
\vspace{-1.5mm}

\noindent\textbf{Exponentially updated shuffler:}
When $f_\T$ and $f_\S$ have the same architecture, we initialize the solver's weights with the pre-trained shuffler's weights, $\bm{\theta}_\S^{(0)}=\bm{\theta}_\T^{(0)}$, and update $\bm{\theta}_\T$ at every epoch end:
\begin{align}
  \label{eqn:teacher_update}
    \bm{\theta}_{\T}^{(j+1)} = \alpha\bm{\theta}_{\T}^{(j)} + (1-\alpha)\bm{\theta}_{\S}^{(j)},
\end{align}
where $\alpha\in[0,1]$ and $j$ is the epoch index.

\noindent\textbf{Frozen shuffler:}
When $f_\T$ is pre-trained with MixIT, we have another option to train $f_\S$ with fewer outputs, e.g., $N_\S=3$.
In such cases, $f_\S$ is trained from scratch without updating $f_\T$.

Note that the sequential update, which replaces $f_\T$ with the latest $f_\S$ and $f_\S$ with a randomly initialized deeper model every several epochs, has been reported to scale better than other protocols and was used as the default strategy in RemixIT~\cite{remixit_journal}.
However, in this work, we focus on the above two protocols to train models efficiently without increasing the parameters and computational costs.

\vspace{-2mm}
\subsection{Relation to previous remixing-based methods}
\label{ssec:prior_work}
\vspace{-1.5mm}

\textbf{RemixIT}:
RemixIT is a self-training framework where a student model $f_\S$ is trained to separate pseudo-mixtures using the outputs of the teacher $f_\T$ as supervision, as described in Sect.~\ref{sssec:in_batch}.
Compared to Self-Remixing, the teacher corresponds to the shuffler and the student to the solver and we use the same protocols described in Sect.~\ref{ssec:shuffler_update} for updating the teacher.
The difference between RemixIT and Self-Remixing lies in the loss computation.
RemixIT uses the teacher's outputs as supervision (Eq.~(\ref{eqn:remixit_loss})) while Self-Remixing uses the observed mixtures (Eq.~(\ref{eqn:proposed_loss})).
It has been shown that if the student sees a large number of pseudo-mixtures from the \textit{same} teacher model, RemixIT loss ideally approaches the supervised loss (see Section II-C in~\cite{remixit_journal}).
Conversely, RemixIT can be sensitive to the teacher model update and frequent updating of the teacher is undesirable since both student's inputs and supervision are made up of the teacher's outputs.
In contrast, Self-Remixing is expected to be more stable because it exploits observed mixtures as supervision, which remain constant even if the shuffler is updated.

\noindent\textbf{RCCL}:
RCCL goes through almost the same process as Self-Remixing, except that it uses the same separator in the shuffler and the solver.
While Self-Remixing only trains the solver via backpropagation, RCCL computes the gradients through two separation and remixing processes.
Thus, the memory cost becomes higher.
In RCCL, we use between-two-mixtures remixing, as described in Sect.~\ref{sssec:between_two_mixtures}.
We confirm that RCCL soon falls into the trivial solution, e.g., $\tilde{s}_{b,1}=x_b, \tilde{s}_{b,{n \neq 1}}=0$ for all $b$, when using in-batch remixing because the model can easily find the solution to minimize the loss.
In contrast, Self-Remixing is less likely to be trapped in the trivial solution and works well even when remixing in a batch because the solver does not see how the shuffler makes pseudo-mixtures.
\vspace{-2.5mm}
\section{Experiments}
\label{sec:setup}
\vspace{-2mm}

\begin{table*}[t]
\begin{center}
\caption{
    Average SISDR [dB], STOI, PESQ, and WER on WSJ-mix test set. 
    $N_\T/N_\S$ are number of shuffler's/solver's separation outputs and $N_\mathcal{R}$ is number of sources used in remixing.
    Memory and duration denote allocated memory and required time for one epoch, respectively.  
}
\label{table:results_wsj}
\resizebox{0.85\linewidth}{!}{
\begin{tabular}{llcccrrcccc}
\toprule
{Method} & {Remix algo.} & {${N_\T}$} & {${N_\S}$} & {${N_\mathcal{R}}$} & {Memory} & {Duration} & {SISDR$^\uparrow$} & {STOI$^\uparrow$} & {PESQ$^\uparrow$} & {WER$^\downarrow$}  \\

\midrule
    Unprocessed &- &- &- &- &- &- &-0.4  &0.726 &1.25 &81.4\% \\ 

\midrule
    \texttt{A0} MixIT   &- &- &6 &- &6.9  GiB &9.2 min &8.7 &0.872 &1.79 &36.5\% \\

\midrule
    \texttt{A1} RCCL    &B/w two mix. &- &6 &6 &14.8 GiB &51.0 min &\textbf{10.0} &0.889 &1.81 &34.4\%   \\
    \texttt{A2} Self-Remixing   &B/w two mix. &6 &6 &6 &8.1 GiB  &50.9 min &\textbf{10.0} &0.889 &1.81 &34.4\%  \\ \hdashline

    \texttt{A3} RemixIT &In batch &6 &6 &3 &7.3 GiB &16.8 min &9.7 &0.887 &1.80 &37.5\%   \\
    \texttt{A4} Self-Remixing &In batch &6 &6 &3 &7.3 GiB &16.8 min &9.9 &\textbf{0.890} &\textbf{1.84} &\textbf{33.6}\%   \\ \hdashline
    
    \texttt{A5} RemixIT &In batch &6 &3 &3 &7.2 GiB &16.0 min &9.6 &0.885 &1.80 &35.6\%   \\
    \texttt{A6} Self-Remixing   &In batch &6 &3 &3 &7.2 GiB &16.0 min &9.5 &0.882 &1.71 &39.5\%   \\

\midrule
    Supervised PIT &- &- &3 &- &6.7 GiB &13.0 min &10.5  &0.908 &2.02 &25.4\% \\

\bottomrule

\end{tabular}}
\end{center}
\vspace{-9mm}
\end{table*}

\vspace{-1.5mm}
\subsection{Datasets}
\label{sec:datasets}
\vspace{-1.5mm}

\textbf{WSJ-mix}:
We synthesized reverberant noisy two-speaker mixtures using speeches from WSJ0~\cite{wsj0} and WSJ1~\cite{wsj1} and noises from CHiME3 dataset~\cite{chime3} sampled at 16~kHz. 
We used almost the same configurations as those in SMS-WSJ~\cite{smswsj} but changed the sampling frequency, the reverberation times, and the noise type and level.
We used Pyroomacoustics~\cite{pra} to simulate RIRs, where the reverberation times were chosen from \SIrange{200}{600}{\milli\second}.
SNR of noises ranged between \SIrange{10}{20}{\decibel}.
Training, validation, and test sets contained 33561 ($\sim$87.4h), 982 ($\sim$2.5h), and 1332 ($\sim$3.6h) mixtures, respectively.
We used the ASR backend provided in~\cite{smswsj} to evaluate WERs.
Since our ASR model was trained on speeches sampled at 8~kHz, we downsampled sources before evaluating WERs.

\noindent\textbf{LibriCSS}:
LibriCSS~\cite{libricss} contained ten hours of recordings obtained by playing back utterances from Librispeech~\cite{librispeech} test set from loudspeakers in a conference room.
Each mixture contained one or two speakers.
LibriCSS consisted of ten sessions, and we used sessions 2-9 for training, session 1 for validation, and session 0 for the test.
We evaluated WERs \textit{utterance-wisely} with the default hybrid system in~\cite{libricss} and Transformer-based model from the ESPNet~\cite{espnet}.\footnote{\url{https://zenodo.org/record/3966501#.Y1f4nuxBz0o}}

\vspace{-4mm}
\subsection{Separation model}
\label{sec:architecture}
\vspace{-1.5mm}

We used Conformer~\cite{conformer} as the separation model (implemented based on~\cite{libricss_conformer}).
The model consisted of 16 Conformer encoder layers with four attention heads, 256 attention dimensions, and 1024 feed-forward network dimensions.
We replaced the batch normalization~\cite{batchnorm} with the group normalization~\cite{groupnorm} with eight groups.
The model took log magnitude spectrograms in the STFT domain as inputs and output time-frequency masks.
The FFT size was 512 with a Hanning window with a size of 400 and a hop length of 160.

\vspace{-4mm}
\subsection{Training and evaluation details}
\vspace{-1.5mm}
As the signal-level loss function $\mathcal{L}$, we used the negative thresholded SNR between the reference $y$ and the estimate $\hat{y}$:
\begin{align}
  \label{eqn:snr_loss}
    \L(y, \hat{y}) = 10\log_{10}{(||y-\hat{y}||^2 + \tau||y||^2)} - 10\log_{10}{||y||^2},
\end{align}
where $\tau = 10^{-3}$ is a soft threshold that clamps the SNR at 30~\si{\decibel}.

We conducted unsupervised learning and semi-supervised domain adaptation.
The batch size was 8 and the input was 6 seconds long.
The model parameters were optimized using the AdamW optimizer~\cite{adamw} with the weight decay of 1e-2.
We linearly increased the learning rate from 0 to the peak learning rate $Lr$ in the first 40000 training steps.
In RemixIT, we did not ensure MC in the student because it rather lowered performance.
In RCCL, we used checkpointing~\cite{checkpointing, dmc} to reduce memory cost.
We set $l_{\rm{thres}}$ to -15 when remixing between two mixtures.
We used $f_\S$ for testing.
\\
\textbf{Unsupervised learning:}
We first trained models with MixIT for 400 epochs and further trained with RCCL, RemixIT, or Self-Remixing for 250 epochs.
WSJ-mix was used in this experiment.
$Lr$ was 2e-4 in MixIT (\texttt{A0} in Table~\ref{table:results_wsj}), 2e-5 when updating $f_\T$ or RCCL (\texttt{A1}-\texttt{A4}), and 5e-4 when $f_\S$ was trained from scratch with the frozen $f_\T$ (\texttt{A5}, \texttt{A6}), and decayed by 0.98 for every two epochs.
The gradients were accumulated for eight training steps.
We enforced MC on \texttt{A4} and \texttt{A6} in inference because such worked better.
When updating $f_\T$, $\alpha$ in Eq.~(\ref{eqn:teacher_update}) was set to 0.8.
For testing, we used the averaged model parameters of the five checkpoints that gave the highest scale-invariant signal-to-distortion ratio (SISDR)~\cite{sisdr} on the validation set.\footnote{Although using in-domain validation data with ground truths is not realistic in the unsupervised setup, we utilized them for a fair comparison.}
Two separated signals with the highest powers were evaluated.

\noindent\textbf{Semi-supervised domain adaptation:}
Models were first trained by supervised PIT on WSJ-mix and then adapted to LibriCSS in a semi-supervised manner.
The models had two output channels for speeches and one for noise.
One mini-batch was composed of four data from WSJ-mix and four from LibriCSS, and the loss was computed by summing up the supervised loss on WSJ-mix and the unsupervised loss on LibriCSS.
We applied gradient accumulation for 16 steps.
$Lr$ was set to 1e-5 and decayed by 0.97 for every epoch, where we defined an \textit{epoch} as the number of steps in which all WSJ-mix data were used once ($33561/4$ steps).
Domain adaptation was conducted for 100 epochs with $\alpha=0.9$.
We tested the averaged model parameters of the five checkpoints that gave the best WERs for the Transformer-based ASR model on the validation set.

\vspace{-3mm}
\begin{figure}[t]
\centering
\centerline{\includegraphics[width=0.85\linewidth]{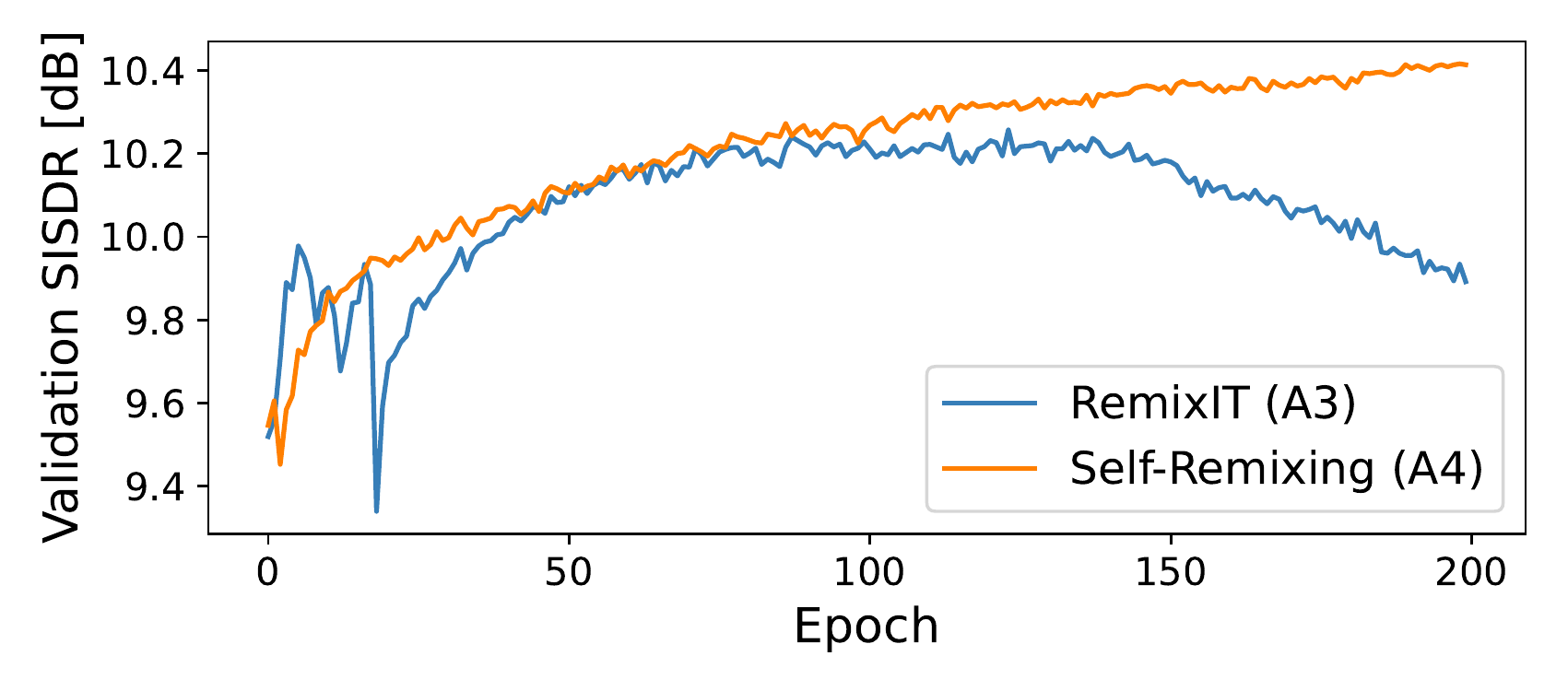}}
\vspace{-6mm}
\caption{
    Validation SISDRs of $f_\S$ of RemixIT and Self-Remixing in first 200 epochs on WSJ-mix.
    Initial performance slightly differs because \texttt{A4} enforces mixture consistency in inference.
}
\label{fig:training_curve}
\vspace{-5mm}
\end{figure}
	
\vspace{-2mm}
\subsection{Results of unsupervised speech separation}
\label{ssec:unsupervised}
\vspace{-1.5mm}

Table \ref{table:results_wsj} lists the SISDR~\cite{fast_bss_eval}, the short-time objective intelligibility (STOI)~\cite{stoi}, the perceptual evaluation of speech quality (PESQ)~\cite{pesq}, and WER on WSJ-mix.
We also measured the allocated memory with \texttt{nvidia-smi} command and the duration for an epoch on NVIDIA GeForce RTX 3090 GPU, without checkpointing.

First of all, Self-Remixing gave the same performance as RCCL when using the same remixing method, with fewer memory costs (\texttt{A1} vs. \texttt{A2}).
Additionally, while RCCL fell into the trivial solution when remixing in a batch (see Sect.~\ref{ssec:prior_work}), Self-Remixing performed well with much less training costs (\texttt{A4}).
The substantial difference in duration between \texttt{A1} and \texttt{A4} is due to the fact that the former needs to compute 400 ($=(C(6,3))^2$) permutations for computing the loss, while the latter computes only 6 ($=3!$) permutations.
These results show the superiority of Self-Remixing in terms of training costs and stability, giving the same or higher performance.

While Self-Remixing outperformed RemixIT when the shuffler/teacher was updated (\texttt{A3} vs. \texttt{A4}), RemixIT performed better when the shuffler/teacher was frozen (\texttt{A5} vs. \texttt{A6}).
It implies that the shuffler updating is essential for Self-Remixing, while RemixIT is more stable when the teacher is frozen since RemixIT essentially depends on the teacher heavily (see Sect.~\ref{ssec:prior_work}).
Fig.~\ref{fig:training_curve} shows the validation SISDR of \texttt{A3} and \texttt{A4} in the first 200 epochs.
The performance of Self-Remixing improves as the training progresses, whereas that of RemixIT drops in the middle of the training.
These results show that Self-Remixing achieves more stable training and higher performance by exploiting observed mixtures as supervisions.

Overall, all the methods improved the MixIT pre-trained model in terms of SISDR and STOI, whereas PESQ and WER did not improve much and even worsened in some cases.
We believe this is because remixed pseudo-mixtures have distortions that are not present in observed signals, and learning with such signals thus leads to models that tend to output distorted signals.
However, the impact of such a mismatch problem can possibly be alleviated by using additional OOD data with ground truths, i.e., semi-supervised learning.

\vspace{-2mm}
\subsection{Results of semi-supervised domain adaptation}
\label{ssec:semi_supervised}
\vspace{-1.5mm}

\begin{table}[t]
    \vspace{-2mm}
	\centering
	\caption{
	    Average WER [\%] on LibriCSS. 
	    Two numbers are WERs of hybrid model in~\cite{libricss} and Transformer-based model from~\cite{espnet}.
    }
	\label{table:results_libricss}
        \resizebox{1\linewidth}{!}{
	\begin{tabular}{lccc}
		\toprule 
		Method &All data & Valid. & Test \\
		
		\midrule
		Unprocessed &26.4 / 20.0 &23.9 / 19.3  &24.4 / 19.3 \\ \midrule
		\texttt{B0} PIT (WSJ-mix) &21.0 / 12.3 &17.9 / 10.8 &19.7 / 11.7 \\ \midrule
		\texttt{B1} RCCL  &17.7 / 9.8 &15.4 / 8.9 &\textbf{16.0} / \textbf{8.6} \\
		\texttt{B2} RemixIT &17.5 / 9.5 &15.5 / 8.6 &16.3 / 8.9 \\
		\texttt{B3} Self-Remixing &\textbf{17.1} / \textbf{9.2} &\textbf{15.0} / 8.5 &16.1 / 8.7 \\	
    	\texttt{B4} RemixIT+Self-Remixing &17.3 / 9.3 &\textbf{15.0} / \textbf{8.3} &\textbf{16.0} / 8.7 \\
		\bottomrule
	\end{tabular}}
	\vspace{-4mm}
\end{table}

Table~\ref{table:results_libricss} shows the average WERs on LibriCSS.
Since the baselines are unsupervised methods and the performances on all data, including training and unseen data, are essential, we evaluated WERs on all, validation, and test data.
\texttt{B1-B4} used PIT pre-trained model \texttt{B0}, where \texttt{B4} was trained with sum of Eq.~(\ref{eqn:remixit_loss}) and Eq.~(\ref{eqn:proposed_loss}).
RCCL used between-two-mixtures remixing, and others used in-batch remixing.

Overall, all the methods stably worked in the semi-supervised setup and significantly improved the performance over the PIT pre-trained model.
These results imply that the mismatch between remixed pseudo-mixtures and observed mixtures can be alleviated by PIT with additional OOD data, and remixing-based methods can help reduce WERs.
Although RCCL gave the best performance on the test set, it performed poorly in the evaluation on all data and the validation set.
Furthermore, it should be noted that RCCL consumes more memory than the other methods and needs heuristic loss thresholding.
In contrast, the proposed methods (\texttt{B3} and \texttt{B4}) performed well, regardless of evaluation sets, with fewer training costs and without the heuristic loss thresholding.
\vspace{-2mm}
\section{Conclusion}
\label{sec:conclusion}
\vspace{-2mm}

We have proposed an unsupervised speech separation method, Self-Remixing. 
The solver module is trained to recover observed mixtures from the pseudo-mixtures generated by the shuffler module, while the shuffler's weights are updated by taking the moving average with the solver's.
The experiments demonstrated the superiority of the proposed method in terms of performance, stability, and efficiency in unsupervised speech separation and semi-supervised domain adaptation.
In the future, we aim to evaluate our method in the separation of real recordings~\cite{ami} or non-speech sounds~\cite{universal_sound_separation}.

\vspace{-2mm}
\section{Acknowledgments}
\label{sec:acknowledgments}
\vspace{-2mm}

The research was supported by NII CRIS collaborative research program operated by NII CRIS and LINE Corporation.

\newpage
\fontsize{8.7pt}{0pt}\selectfont
\bibliographystyle{IEEEbib}
\bibliography{main}

\end{document}